\documentclass[aps,pre,twocolumn,groupedaddress,showkeys]{revtex4}
\usepackage{graphics,color}

\usepackage[]{graphicx}
\DeclareGraphicsExtensions{.pdf}

\begin{document}

\title{The simplest maximum entropy model for collective behavior in a neural network}

\author{Gasper Tka\v{c}ik,$^a$  Olivier Marre,$^{b,e}$ Thierry Mora,$^c$ Dario Amodei,$^{d,f}$ Michael J Berry II,$^{e,f}$ and William Bialek,$^{d,g}$}
\affiliation{$^a$Institute of Science and Technology Austria, Am Campus 1, A--3400 Klosterneuburg, Austria,\\
$^b$Institution de al Vision, UMRS 968 UPMC, INSERM, CNRS U7210, CHNO Quinze--Vingts, F--75012 Paris, France,\\
$^c$Laboratoire de Physique Statistique de l'\'Ecole Normale Superieure, CNRS and Universites Paris VI and Paris VII, 24 rue Lhomond, 75231 Paris Cedex 05, France,\\
$^d$Joseph Henry Laboratories of Physics, $^e$Department of Molecular Biology,  $^f$Princeton Neuroscience Institute, and
$^g$Lewis--Sigler Institute for Integrative Genomics\\
Princeton University, Princeton, New Jersey 08544 USA}

\date{\today}

\begin{abstract}
Recent work emphasizes that the maximum entropy principle provides a bridge between statistical mechanics models for collective behavior in neural networks and experiments on networks of real neurons. Most of this work has focused on capturing the measured correlations among pairs of neurons. Here we suggest an alternative, constructing models that are consistent with the distribution of global network activity, i.e. the probability that $K$ out of $N$ cells in the network generate action potentials in the same small time bin. The inverse problem that we need to solve in constructing the model is analytically tractable, and provides a natural ``thermodynamics'' for the network in the limit of large $N$. We analyze the responses of neurons in a small patch of the retina to naturalistic stimuli, and find that the implied thermodynamics is very close to an unusual critical point, in which the entropy (in proper units) is exactly equal to the energy.
\end{abstract}

\maketitle

Many of the most interesting phenomena of life are collective, emerging from interactions among many elements, and physicists have long hoped that these collective biological phenomena could be described within the framework of statistical mechanics. One approach to a statistical mechanics of biological systems is exemplified by Hopfield's discussion of neural networks, in which simplifying assumptions about the underlying dynamics lead to an effective ``energy landscape'' on the space of network states \cite{hopfield_82,amit_89,hertz+al_91}. In a similar spirit, Toner and Tu showed that simple stochastic dynamical models for coordinating the motion of moving organisms, as in flocks of birds or schools of fish, can be mapped to an effective field theory in the hydrodynamic limit \cite{toner+tu_95,toner+tu_98}.

A very different way of constructing a statistical mechanics for real biological systems is through the maximum entropy principle \cite{jaynes_57}. Rather than making specific assumptions about the underlying dynamics, we take a relatively small set of measurements on the system as given, and build a model for the distribution over system states that is consistent with these experimental results but otherwise has as little structure as possible. This automatically generates a Boltzmann--like distribution, defining an energy landscape over the states of the system; importantly, this energy function has no free parameters, but is completely determined by the experimental measurements. As an example, if we look in small windows of time where each neuron in a network either generates an action potential (spike) or remains silent, then the maximum entropy distribution consistent with the mean probability of spiking in each neuron and the correlations among spikes in pairs of neurons is exactly an Ising spin glass \cite{schneidman+al_06}. Similarly, the maximum entropy model consistent with the average correlations between the flight direction of a single bird and its immediate neighbors in a flock is a Heisenberg model \cite{bialek+al_12}. Starting with the initial work on the use of pairwise maximum entropy models to describe small ($N=10 - 15$) networks of neurons in the retina, this approach has been used to describe the activity in a variety of neural networks \cite{shlens+al_06,tkacik+al_06,yu+al_08,tang+al_08,tkacik+al_09,shlens+al_09,ohiorhenuan+al_10,ganmor+al_11}, the structure and activity of biochemical and genetic networks \cite{lezon+al_06,tkacik_07},  the statistics of amino acid substitutions in protein families 
\cite{bialek+ranganathan_07,seno+al_08,weigt+al_09,halabi+al_09,mora+al_10,marks+al_11,sulkowska+al_12}, and the rules of spelling in English words \cite{stephens+bialek_10}. Here we return to the retina, taking advantage of new electrode arrays that make it possible to record from a large fraction of the  $\sim 200$ output neurons within a small, highly interconnected patch of the circuitry \cite{expts}. Our goal is not to give a precise model, but rather to construct the simplest model that gives us a glimpse of the collective behavior in this system.  For a different approach to simplification, see Ref \cite{macke+al_11}.

The maximum entropy approach is much more general than the construction of models based on pairwise correlations. To be concrete, we consider small slices of time during which each neuron in our network either generates an action potential or remains silent. Then the states of individual neurons are defined by $\sigma_{\rm i} =1$  when neuron $\rm i$ generates a spike, and $\sigma_{\rm i} =-1$ when neuron $\rm i$ is silent. States of the entire network are defined by $\vec\sigma \equiv \{\sigma_{\rm i}\}$, and we are interested in the probability distribution of these states, $P(\vec \sigma )$. If we know the average values of some functions $f_\mu (\vec\sigma )$, then the maximum entropy distribution consistent with this knowledge is
\begin{equation}
P(\vec\sigma ) = {1\over{Z(\{g_\mu\})}} \exp\left[ - \sum_\mu g_\mu f_\mu (\vec\sigma ) \right], 
\end{equation}
where the couplings $g_\mu$ have to be adjusted to match the measured expectation values $\langle f_\mu (\vec\sigma )\rangle$.

In any given slice of time, we will find that $K$ out of the $N$ neurons generate spikes, where
\begin{equation}
K = {1\over 2}\sum_{{\rm i}=1}^N (\sigma_{\rm i} + 1).
\end{equation}
One of the basic characteristics of a network is the distribution of this global activity, $P_N(K)$. As an example, in Fig 1 we show experimental results on $P_N(K)$ for groups of $N = 40$ neurons in the retina as it views a naturalistic movie. In these experiments \cite{expts}, we use a dense array of electrodes that samples $160$ out of the $\sim 200$ ganglion cells in a small patch of the salamander retina, and we divide time into bins of $\Delta\tau  = 20\,{\rm ms}$. The figure shows the average behavior in groups of $N = 40$ cells chosen out of this network, under conditions where a naturalistic movie is projected onto the retina. The correlations between pairs of cells are small, but $P_N(K)$ departs dramatically from what would be expected if the neurons generated spikes independently.

\begin{figure}
\includegraphics[width=\linewidth]{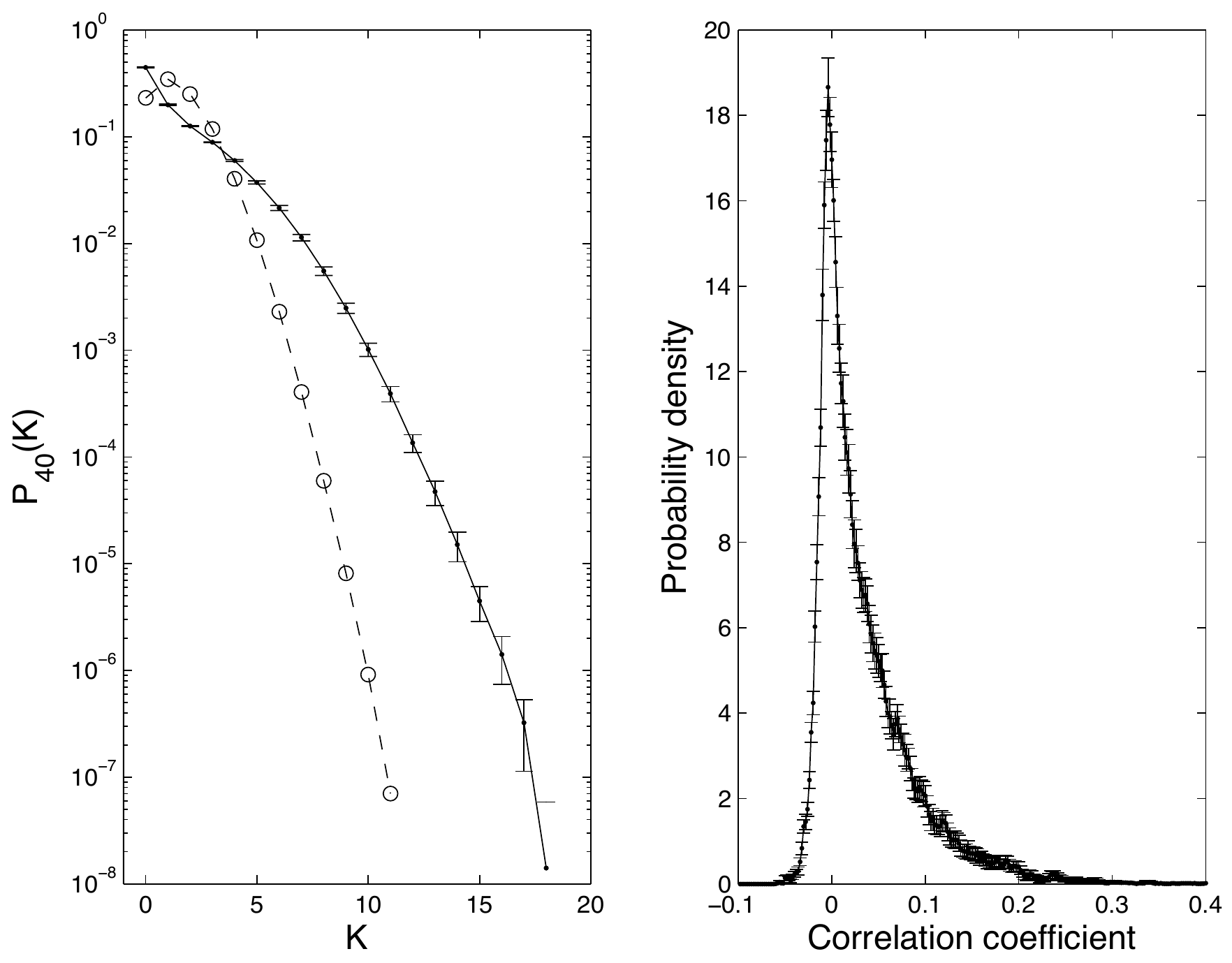}
\caption{Experimental results for $P_N (K)$, in groups of $N = 40$ neurons. At left, solid points show the distribution estimated by averaging over many randomly chosen groups of $N = 40$ cells out of the $N = 160$ in our data set; error bars are standard deviations across random halves of the duration of the experiment. Open circles are the expectation if cells are independent. At right, the distribution of correlation coefficients among pairs of neurons in out sample. Because the experiment is long, the threshold for statistical significance of the correlations is very low, $|C_{\rm thresh}| \leq 0.01$. Almost all pairs of cells thus have significant correlations, but these correlations are weak.}
\end{figure}

How do we construct the maximum entropy model consistent	with the	measured	$P_N (K)$? 
Knowing the distribution $P_N (K )$ is equivalent to knowing	all	its	moments, so the functions $f_\mu(\vec \sigma)$ whose expectation values we have measured are
$f_1(\vec\sigma ) = K$,
$f_2 (\vec\sigma ) = K^2$,
and so on. Thus we can write
\begin{equation}
P_N(\vec\sigma ) = {1\over{Z(\{g_\mu\})}} \exp\left[-\sum_{n=1}^N g_n K^n\right]
= {1\over {Z_N}} e^{-V_N(K)} ,
\label{PNdef}
\end{equation}
where $V_N(K)$ is some effective potential that we need to choose so that $P_N (K)$ comes out equal to the experimentally measured	$P_N^{\rm exp} (K)$.

Usually the inverse problem for these maximum entropy distributions is hard. Here it is much easier. We note that
\begin{eqnarray}
P_N(K) &\equiv& \sum_{\vec \sigma} \delta\left[ K, {1\over 2}\sum_{{\rm i}=1}^N (\sigma_{\rm i}+1)\right] P(\vec\sigma )\\
&=& {1\over {Z_N}}{\cal N}(K,N) e^{-V_N (K)} ,
\end{eqnarray}
where 
\begin{equation}
{\cal N}(K,N) = {{N!}/{(N-K)!K!}}. 
\label{SNK}
\end{equation}
The log of this number is an entropy at fixed $K$, $S_N(K) \equiv \ln {\cal N}(K,N)$, so we can write
\begin{equation}
P_N(K) = {1\over{Z_N}} \exp\left[ S_N(K) - V_N(K)\right] .
\end{equation}
Finally, to match the distribution $P_N(K)$ to the experimental measurement $P_N^{\rm exp}(K)$, we must have
\begin{equation}
V_N(K) = -\ln P_N^{\rm exp}(K) +S_N(K) + \ln Z_N .
\end{equation}
In Fig 2 we show the average results for $V_N(K)$ in networks of size $N = 40$.

We expect that both energy and entropy will be extensive quantities.  For the entropy $S_N(K)$ this is guaranteed by Eq (\ref{SNK}), which tells us that as $N$ becomes large, $S_N(K)  \rightarrow N s(K/N )$.	It is an experimental	question	whether,	in the networks we are studying, there is something analogous to a thermodynamic limit in which, for large $N$, we have $V_N(K) \rightarrow N\epsilon(K/N)$. This is illustrated at right in Fig 2, where for $K/N = 0.05$ we study the dependence of the energy per neuron on $1/N$. There is a natural extrapolation to large $N$, and this is true for all the ratios of $K/N$ that we tested.

\begin{figure}[thb]
\includegraphics[width=\linewidth]{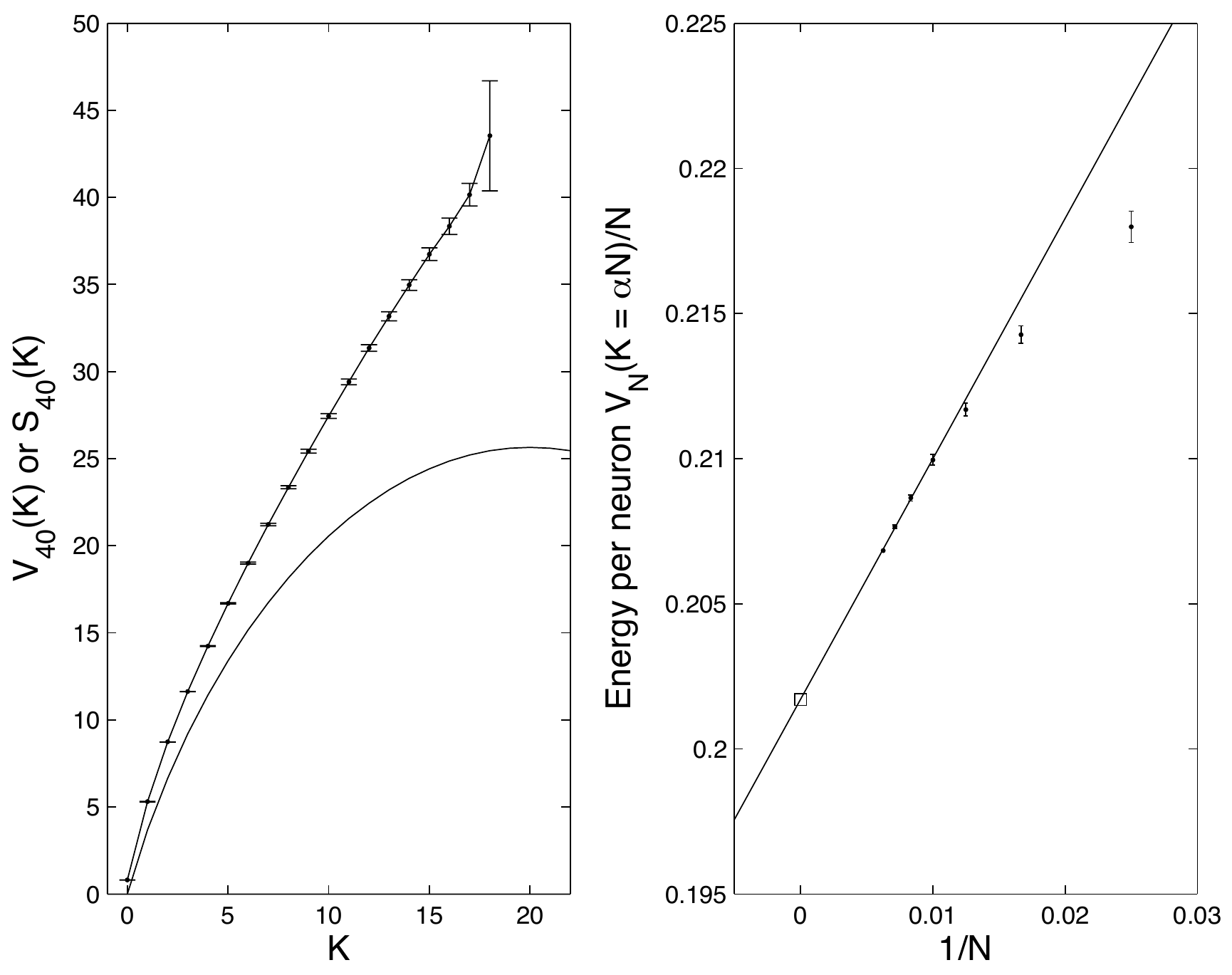}
\caption{The effective potential and its dependence on system size. At left, results for $N = 40$ neurons, showing both the potential $V_N(K)$ (points with error bars) and the entropy $S_N (K)$ (smooth curve); error bars are as in Fig 1. At right, the behavior of $V_N (K = \alpha N )/N$, for $\alpha = 0.05$, showing the dependence on $N$ (points with error bars) and the extrapolation   $N\rightarrow \infty$ (square).}
\end{figure}

In the $N \rightarrow \infty$ limit, the natural quantities are the energy and entropy per neuron, $\epsilon$ and $s$, respectively, and these are shown in Fig 3. One clear result is that, as we look at more and more neurons in the same patch of the retina, we do see the emergence of a well defined, smooth relationship between entropy and energy $s(\epsilon )$. While most neural network models are constructed so that this thermodynamic limit exists, it is not so obvious that this should happen in real data. In particular, if we consider a family of models with varying $N$ in which all pairs of neurons are coupled, the standard way of arriving at a thermodynamic limit is to scale the coupling strengths with $N$, and correspondingly the pairwise correlations are expected to vary with $N$. In constructing maximum entropy models, we can't follow this path, since the correlations are measured and thus by definition don't vary as we include more and more neurons. Here we focus not on correlations but on the distribution $P_N(K)$, and thus the emergence of a thermodynamic limit depends on the evolution of this distribution with $N$.

\begin{figure}
\includegraphics[width=\linewidth]{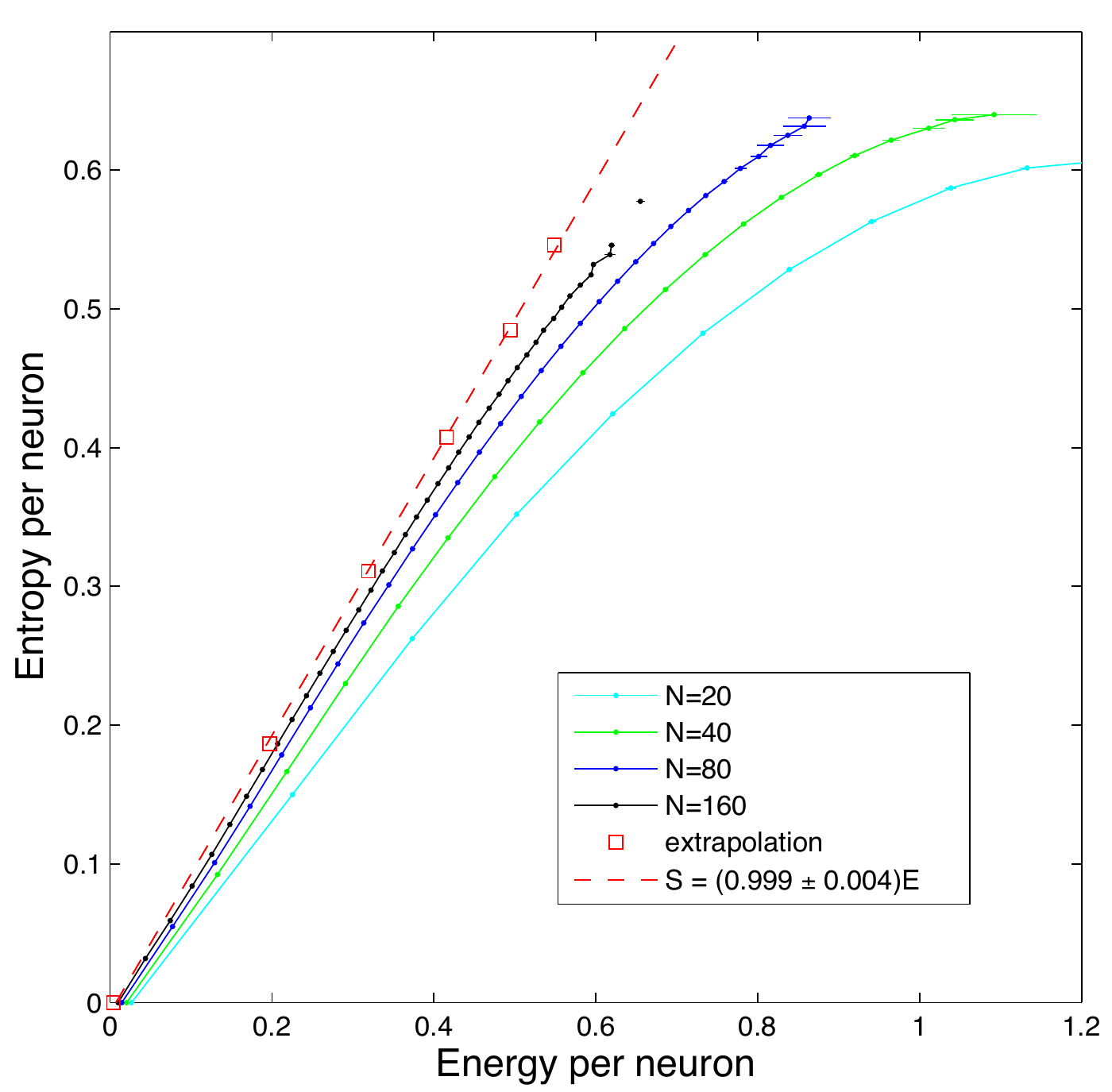}
\caption{Entropy vs. energy. We compute the effective energy per neuron,  $\epsilon = V_N(K)/N$,  averaged over multiple groups of $N$ neurons chosen out of the $160$ we have access to in the experiment, and then compare with the entropy per neuron,  $s = S_N(K)/N$.  The extrapolation is as in Fig 2, and error bars in energy (visible only when larger than symbols) are as in Fig. 1.}
\end{figure}

We recall that the plot of entropy vs. energy tells us everything about the thermodynamics of the system. In our maximum entropy construction, there is no real temperature---$k_B T$ just provides units for the effective energy $V_N (K)$. But if we have a model for the energy as a function of the microscopic state of the system, then we can take this seriously as a statistical mechanics problem and imagine varying the temperature. More precisely, we can generalize Eq (\ref{PNdef}) to consider
\begin{equation}
P_N(\vec\sigma  ; \beta ) = {1\over {Z_N(\beta )}} e^{-\beta V_N(K)} ,
\end{equation}
where the real system is at $\beta = 1$.  Then in the thermodynamic limit we have the usual identities: the temperature is defined by $\partial s/\partial \epsilon = \beta$, the specific heat is $C=k_B \beta^2 (-\partial^2 s /\partial\epsilon^2)^{-1}$, and so on.   In particular, the vanishing of the second derivative of the entropy implies a diverging specific heat, a signature of a critical point.

In our case, since the real system is at $\beta = 1$, the behavior of the network will be dominated by states with an energy per neuron that solves the equation $\partial s/\partial \epsilon = 1$. But Fig 3 shows us that, as we consider more and more neurons, the function $s(\epsilon )$ seems to be approaching $s = \beta_0 \epsilon$, where $\beta_0 = 0.999\pm 0.004$ is  one within errors. If we had exactly $s = \epsilon$, then all energies would be solutions of the condition $\partial s/\partial \epsilon = 1$. Correspondingly, the specific heat $C$ would diverge, signaling that $\beta = 1$ is a critical point. This is a very unusual critical point, since all higher derivatives of the entropy vanish \cite{mora+bialek_11}.

\begin{figure}[tb]
\includegraphics[width=\linewidth]{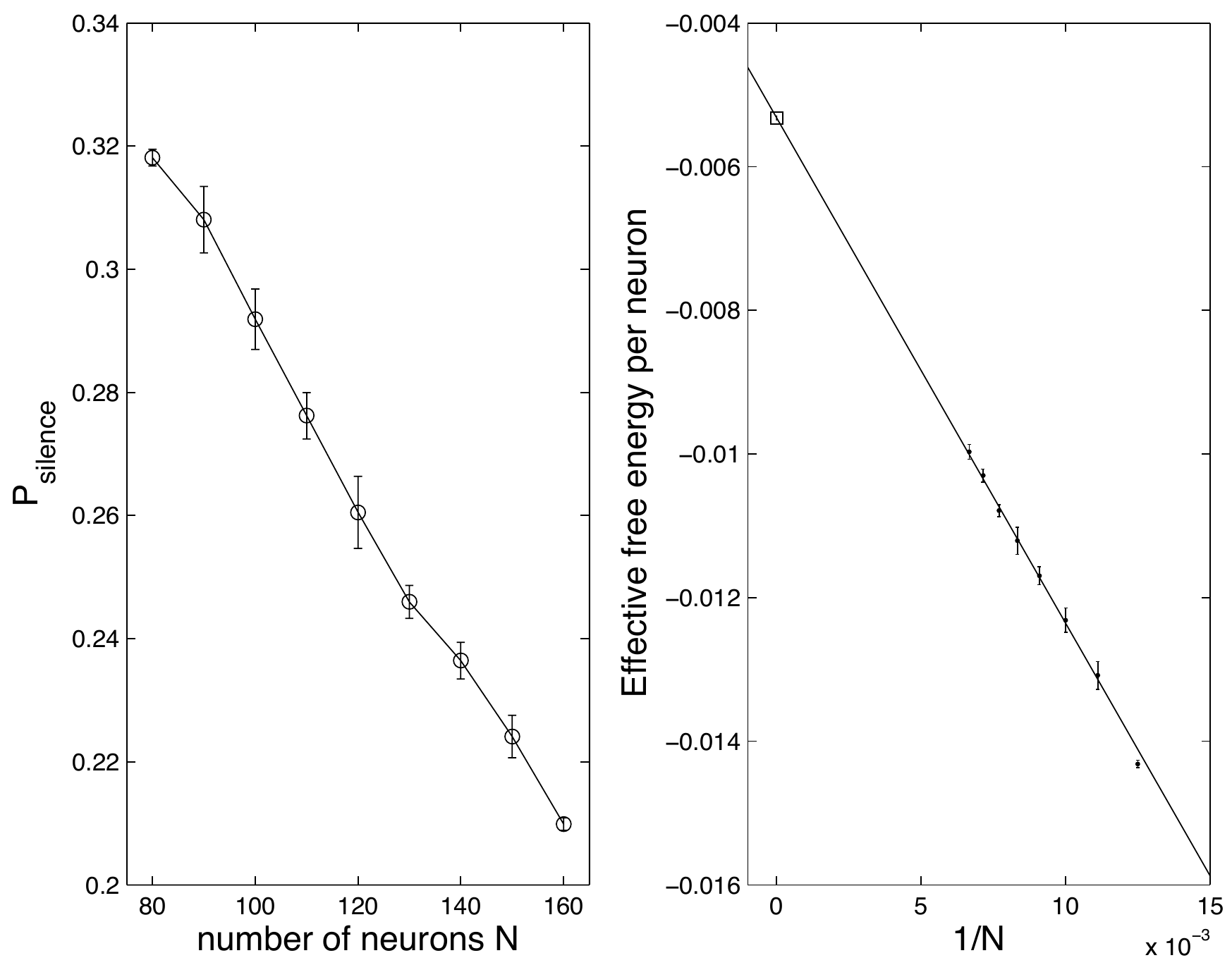}
\caption{The probability of silence, and the effective free energy. At left, the probability that a network of $N$ neurons is in the silent state, where none of the cells generate a spike within a window $\Delta\tau$; error bars as in Fig 1. Note that this probability declines very slowly with the number of neurons $N$. At right, we translate the probability of silence into an effective free energy per neuron, and see that this varies linearly with $1/N$, the extrapolation $N\rightarrow \infty$ (square).}
\end{figure}

More generally, when we try to describe the probability distribution over states $\vec\sigma$ using ideas from statistical mechanics, we are free to choose the zero of the (effective) energy as we wish. A convenient choice is that the unique state of zero spikes---complete silence in the network---should have zero energy.   Unless there are exponentially many states that with probability equal to the silent state (which seems unlikely), in the large $N$ limit the entropy per neuron will also be zero at zero energy.  But with this choice for the zero of energy, the probability of the silent  state is given by $P_{\rm silence} = 1/Z$, and $Z = e^{-F}$, where $F$ is the effective free energy, since we are at $\beta = 1$.  Thus if we can measure this probability reliably, we can ``measure'' the free energy, without any further assumptions.   We see in Fig 4 that the probability of silence falls as we look at more and more neurons, which makes sense since the free energy should grow with system size. But the decline in the probability of silence is surprisingly slow. We can make this more precise by computing the effective free energy per neuron, $f = F/N$, also shown. This is a very small number indeed, $f \sim  -0.01$ at the largest values of $N = 160$ for which we have data. 

We recall that, with $k_B T = 1$,  the free energy per neuron is $f = \langle \epsilon\rangle - s_{\rm total}$, where $\langle \epsilon\rangle$ denotes the average energy and $s_{\rm total}$ is the total entropy of the system, again normalized per neuron.  Our best estimate of the entropy of the states taken on by the network is $s_{\rm total} \sim 0.2$ per neuron, which means that the free energy reflects a cancellation between energy and entropy with a precision of at least $\sim  5\%$. If we extrapolate to the thermodynamic limit the cancellation becomes even more precise, so that the extensive component of the free energy is $f_\infty = -0.0051 \pm  0.00003$ (Fig 4). Notice that the small value of the free energy means that the silent state occurs frequently, and hence we can measure its probability very accurately, so the error bars are small. If we had a critical system in which $s(\epsilon ) = \epsilon$, the extensive component of the free energy would be exactly zero. 

In a normal thermodynamic limit (and $\beta = 1$), $f_\infty = \epsilon^* - s(\epsilon^*)$, where $\epsilon^*$ is the energy at which $\partial s/\partial \epsilon = 1$.  Geometrically, $f_\infty$ is the intercept along the energy axis of a line with unit slope that is tangent to the curve $s(\epsilon )$ at the point $\epsilon^*$.    From above we have $s(0) = 0$, and then if $s(\epsilon )$ is concave ($\partial^2 s(\epsilon ) /\partial\epsilon^2 < 0$, so that the specific heat is everywhere positive) we are guaranteed that $f_{\infty}$ is negative.  But to have $f_{\infty}\rightarrow 0$ then requires that $\partial s(\epsilon )/\partial \epsilon \leq 1$ at $\epsilon = 0$.   In this scenario, pushing $f_\infty$ toward zero requires both $\epsilon^*$ and $s(\epsilon^*)$ to approach zero, so that the network is in a (near) zero entropy state despite the finite temperature.  This state would be similar to the critical point in the random energy model \cite{rem}, but this seems inconsistent with the evidence for a nonzero entropy per neuron.  

To have near zero free energy with nonzero entropy seems to require something very special.  One possibility is to allow  $\partial^2 s(\epsilon ) /\partial\epsilon^2 > 0$, allowing phase coexistence between the $\epsilon = 0$ silent state and some other $\epsilon\neq 0$ state.  The other possibility is to have $s(\epsilon ) = \epsilon$, as suggested by Fig 3.  Thus, while  the observation of a nearly zero free energy per neuron does not prove that the entropy is equal to the energy for all energies, it does tell us that the network is in or near one of a handful of unusual collective states.

The model we have considered here of course throws away many things: we are not keeping track of the identities of the cells, but rather trying to capture the global activity of the network. On the other hand, because we are considering a maximum entropy model, we know that what we are constructing is the least structured model that is consistent with $P_N (K)$. It thus is surprising that this minimal model is so singular. As we have emphasized, even without appealing to a model, we know that there is something special about these networks of neurons because they exhibit an almost perfect cancellation of energy and entropy. The more detailed maximum entropy analysis suggests that cancellation is not just true on average, but rather that the entropy is almost precisely equal to the energy as a function. This is consistent with hints of criticality in previous analyses, which extrapolated from much smaller groups of neurons \cite{tkacik+al_06,tkacik+al_09,mora+bialek_11}, although much more remains to be done.

We thank A Cavagna, I Giardina, E Schneidman, GJ Stephens, T Taillefumier, and A Walczak for helpful discussions. This work was supported in part by NSF Grants IIS--0613435 and PHY--0957573, by NIH Grants R01 EY14196 and P50 GM071508, by the Fannie and John Hertz Foundation, by the Human Frontiers Science Program, by the Swartz Foundation, and by the WM Keck Foundation.


\begin{thebibliography}{99}
\bibitem{hopfield_82}
JJ Hopfield, 
Neural networks and physical systems with emergent collective computational abilities. 
{\em Proc Natl Acad Sci (USA)} {\bf 79,} 2554--2558 (1982).
\bibitem{amit_89}
DJ Amit, {\em Modeling Brain Function: The World of Attractor Neural Networks} (Cambridge University Press, Cambridge, 1989).
\bibitem{hertz+al_91}
J Hertz, A Krogh \& RG Palmer, {\em Introduction to the Theory of Neural Computation} (Addison Wesley, Redwood City, 1991).
\bibitem{toner+tu_95}
J Toner \& Y Tu, 
Long--range order in a two--dimensional XY model: How birds fly together. 
{\em Phys Rev Lett} {\bf 75,} 4326--4329 (1995).
\bibitem{toner+tu_98}
J Toner \& Y Tu, 
Flocks, herds, and schools: A quantitative theory of flocking. 
{\em Phys Rev E} {\bf 58,} 4828--4858 (1998).
\bibitem{jaynes_57}
ET Jaynes, 
Information theory and statistical mechanics. 
{\em Phys Rev} {\bf 106,} 620--630 (1957).
\bibitem{schneidman+al_06}
E Schneidman, MJ Berry II, R Segev \& W Bialek. 
Weak pairwise correlations imply strongly correlated network states in a neural population. 
{\em Nature} {\bf 440,} 1007--1012 (2006). 
\bibitem{bialek+al_12}
W Bialek, A Cavagna, I Giardina, T Mora, E Silvestri, M Viale \& A Walczak, 
Statistical mechanics for natural flocks of birds. 
{\em Proc Natl Acad Sci (USA)} {\bf 109,} 4786--4791 (2012); arXiv.org:1107.0604 [physics.bio--ph] (2011).
\bibitem{shlens+al_06}
J Shlens, GD Field, JL Gaulthier, MI Grivich, D Petrusca, A Sher, AM Litke \& EJ Chichilnisky, 
The structure of multi--neuron firing patterns in primate retina. 
{\em J Neurosci} {\bf 26,} 8254--8266 (2006).
\bibitem{tkacik+al_06}
G Tka\v{c}ik, E Schneidman, MJ Berry II \& W Bialek, 
Ising models for networks of real neurons. 
arXiv:q-bio/0611072 (2006).
\bibitem{yu+al_08}
S Yu, D Huang, W Singer \& D Nikolic, 
A small world of neuronal synchrony. 
{\em Cereb Cortex} {\bf 18,} 2891--2901 (2008). 
\bibitem{tang+al_08}
A Tang, D Jackson, J Hobbs, W Chen, JL Smith, H Patel, A Prieto, D Petruscam MI Grivich, A Sher, P Hottowy, W Dabrowski, AM Litke \& JM Beggs, 
A maximum entropy model applied to spatial and temporal correlations from cortical networks in vitro. 
{\em J Neurosci} {\bf 28,} 505--518 (2008). 
\bibitem{tkacik+al_09}
G Tka\v{c}ik, E Schneidman, MJ Berry II \& W Bialek, 
Spin--glass models for a network of real neurons.
arXiv:0912.5409 (2009). 
\bibitem{shlens+al_09}
J Shlens, GD Field, JL Gaulthier, M Greschner, A Sher, AM Litke \& EJ Chichilnisky, 
The structure of large--scale synchronized firing in primate retina. 
{\em J Neurosci} {\bf 29,} 5022--5031 (2009).
\bibitem{ohiorhenuan+al_10}
IE Ohiorhenuan, F Mechler, KP Purpura, AM Schmid, Q Hu \& JD Victor, 
Sparse coding and higher--order correlations in fine--scale cortical networks. 
{\em Nature} {\bf 466,} 617--621 (2010).
\bibitem{ganmor+al_11}
E Ganmor, R Segev \& E Schniedman,
Sparse low--order interaction network underlies a highly correlated and learnable neural population code,
{\em Proc Natl Acad Sci (USA)} {\bf 108,} 9679--9684 (2011).
\bibitem{lezon+al_06}
 TR Lezon, JR Banavar, M Cieplak, A Maritan \& NV Federoff, 
 Using the principle of entropy maximization to infer genetic interaction networks from gene expression patterns. 
 {\em Proc Natl Acad Sci (USA)} {\bf 103,} 19033--19038 (2006).
\bibitem{tkacik_07}
 G Tka\v{c}ik, {\em Information Flow in Biological Networks} (Dissertation, Princeton University, 2007).
\bibitem{bialek+ranganathan_07} W Bialek \& R Ranganathan, 
Rediscovering the power of pairwise interactions. 
arXiv:0712.4397 [q--bio.QM] (2007).
\bibitem{seno+al_08}
F Seno, A Trovato, JR Banavar \& A Maritan, 
Maximum entropy approach for deducing amino acid interactions in proteins. 
{\em Phys Rev Lett} {\bf 100,} 078102 (2008).
\bibitem{weigt+al_09}
 M Weigt, RA White, H Szurmant, JA Hoch \& T Hwa, 
 Identification of direct residue contacts in protein--protein interaction by message passing. 
 {\em Proc Natl Acad Sci (USA)} {\bf 106,} 67--72 (2009).
\bibitem{halabi+al_09}
 N Halabi, O Rivoire, S Leibler \& R Ranganathan, 
 Protein sectors: Evolutionary units of three--dimensional structure. 
 {\em Cell} {\bf 138,} 774--786 (2009).
 \bibitem{mora+al_10}
 T Mora, AM Walczak, W Bialek \& CG Callan, 
 Maximum entropy models for antibody diversity. 
 {\em Proc Natl Acad Sci (USA)} {\bf 107,} 5405--5410 (2010); arXiv:0912.5175 [q--bio.GN] (2009).
 \bibitem{marks+al_11}
 DS Marks, LJ Colwell, R Sheridan, TA Hopf, A Pagnani, R Zecchina \& C Sander, 
 Protein 3D structure computed from evolutionary sequence variation. 
 {\em PLoS One} {\bf 6,} e28766 (2011).
\bibitem{sulkowska+al_12}
 JI Sulkowska, F Morocos, M Weigt, T Hwa \& JN Onuchic, 
 Genomics--aided structure prediction. 
 {\em Proc Natl Acad Sci (USA)} {\bf 109,} 10340--10345 (2012).
\bibitem{stephens+bialek_10}
 GJ Stephens \& W Bialek, 
 Statistical mechanics of letters in words. 
 {\em Phys Rev E} {\bf 81,} 066119 (2010); arXiv:0801.0253 [q--bio.NC] (2008).
\bibitem{expts}
A full account of the experiments will be given elsewhere.    Briefly, experiments were performed on the larval tiger salamander, {\em Ambystoma tigrinum tigrinum}, in accordance with institutional animal care standards. Retinae were isolated from the eye in darkness \cite{puchalla+al_05}, and the retina was pressed, ganglion cell down, against a custom fabricated array of $252$ electrodes (size $8\,\mu{\rm m}$,  spacing $30\,\mu{\rm m}$) \cite{amodei_11}. The retina was superfused with oxygenated Ringer's medium (95\% $\mathrm{O_2}$, 5\% $\mathrm{CO_2}$) at   $22\,^\circ{\rm C}$. Electrode voltage signals were acquired and digitized at 10 kHz by a 252 channel preamplifier (Multi--Channel Systems, Germany). The sorting of these signals into action potentials from individual neurons was done using the methods of Refs \cite{segev+al_04,marre+al_12}.   The stimulus was a $19\,{\rm s}$ grayscale movie clip of a swimming fish and water plants in a fish tank, which was repeated 297 times. It was presented using a CRT display (refresh rate 60 Hz), and focused on the photoreceptor layer of the retina using standard optics.
 \bibitem{macke+al_11}
 JH Macke, M Opper \& M Bethge,
 Common input explains higher--order correlations and entropy in a simple model of neural population activity.
 {\em Phys Rev Lett} {\bf 106,} 208102 (2011).
 \bibitem{rem}
 B Derrida, Random-energy model: An exactly solvable model of disordered systems.  {\em Phys Rev B} {\bf 24,} 2613--2626 (1981).
 \bibitem{mora+bialek_11}
T Mora \& W Bialek, 
Are biological systems poised at criticality?	
{\em J Stat Phys} {\bf 144,} 268--302 (2011);  arXiv:1012.2242 [q--bio.QM] (2010).
\bibitem{puchalla+al_05}
JL Puchalla, E Schneidman, RA Harris \& MJ Berry II, Redundancy in the population code of the retina.  {\em Neuron} {\bf 46,} 493--504 (2005).
\bibitem{amodei_11}
D Amodei, {\em Network--scale Electrophysiology:  Measuring and Understanding the Collective Behavior of Neural Circuits} 
(Dissertation, Princeton University, 2011).
\bibitem{segev+al_04}
R Segev, J Goodhouse, J Puchalla \& MJ Berry II, Recording spikes from a large fraction of the ganglion cells in a retinal patch.  {\em Nature Neurosci} {\bf 7,} 1155--1162 (2004).
\bibitem{marre+al_12}
O Marre, D Amodei, K Sadeghi, F Soo, TE Holy \& MJ Berry II, Recording from a complete population in the retina.  Submitted (2012).
\end{thebibliography}
\end{document}